\documentclass[twocolumn]{aa}
\usepackage{graphicx}
\usepackage{txfonts}
\def\spose#1{\hbox to 0pt{#1\hss}}
\def\simlt{\mathrel{\spose{\lower 3pt\hbox{$\mathchar"218$}}
     \raise 2.0pt\hbox{$\mathchar"13C$}}}
\def\simgt{\mathrel{\spose{\lower 3pt\hbox{$\mathchar"218$}}
     \raise 2.0pt\hbox{$\mathchar"13E$}}}

\def\hi{H\,{\sc i}~}

\begin{document}
   \title{The refill of superbubble cavities}

   \author{S. Recchi
          \inst{1}
          \and
          G. Hensler\inst{1}
          }

   \offprints{S. Recchi}

   \institute{Institute of Astronomy, Vienna University,
              T\"urkenschanzstrasse 17, A-1180 Vienna\\
              \email{recchi, hensler@astro.univie.ac.at}
             }

   \date{Received; accepted}

   \abstract{ In this paper we study the late evolution of model
     galaxies after a single episode of star formation of different
     durations.  The aim of the paper is to discover the timescale
     needed to refill with cold gas the center of the galaxy.  This
     timescale strongly depends on the amount of gas initially
     present inside the galaxy and ranges between 125 and 600 Myr.  A
     \hi hole can therefore survive several hundred Myrs after the
     last SNII has exploded.  If, as a consequence of the refill of
     the center of the galaxy, a second episode of star formation
     occurs, it pollutes the surrounding medium in a very short
     timescale (of the order of 10--15 Myr), at variance with what
     happens if the center of the galaxy is still occupied by hot and
     tenuous gas.

   }

   \maketitle
%
%________________________________________________________________

\section{Introduction}

After an episode of star formation (hereafter SF), the energy released
by SN explosions and stellar winds produces large cavities of hot gas
(visible in \hi as holes), surrounded by a cold \hi shell.  However,
in many dwarf irregular galaxies (hereafter dIrr) there are \hi holes
not associated with young star clusters (e.g. M101, Kamphuis et al.
1991; Holmberg II, Rhode et al. 1999; LMC, Kim et al. 1999).  

The energy input rate after an episode of SF declines with time,
therefore the hot cavity looses pressure and the shell tends to recede
towards the center of the SF region.  This refill process strongly
depends on the considered energy sources.  Superbubbles solely
produced by SNeII experience a buoyancy of the hot gas at the end of
the SF phase and the central region can be completely refilled with
cold gas.  This process occurs in a timescale of the order of a few
10$^8$ yr (D'Ercole \& Brighenti 1999, hereafter DB99).  Also Hunter
\& Gallagher (1990) speculated on the final fate of a supershell,
based on simple dynamical arguments and estimated a refill
timescale of the order of $\sim$ 5 $\times$ 10$^7$ yr.  The assumption
of an energy source coming also from SNeIa (which should indeed be
numerous in irregular galaxies; see e.g.  Mannucci et al.  2005)
changes the thermodynamical behaviour of the gas in the center of
galaxies.  For instance, Recchi et al. (2001) found that in some cases
the break-out of the galaxy occurred only as a consequence of the
energy input from SNeIa, while the SNeII, exploding in a colder and
denser medium, could radiatively loose a very large fraction of the
initial explosion energy.

In some cases, the overpressurized regions produced by the ongoing SF
develop galactic winds.  Galactic winds are defined as
outwards-directed flows of gas, whose expansion velocities reach the
escape velocity, therefore supposed to definitely leave the parent
galaxy.  In stratified media (i.e. if the density profile in the
vertical direction is $\rho_z \propto z^{-x}$), a galactic wind
develops for profiles steeper than x=2 (Koo \& McKee 1992).  However,
the final fate of the outflowing gas depends on poorly constrained
details about the environment surrounding the galaxy and about
interactions with other galaxies and it is hard to distinguish, both
theoretically and observationally, galactic winds from normal
outflows.  However, the refill of a superbubble is not affected at all
by the final fate of the outflowing gas, therefore this distinction is
not important in our work.

The refill of the center of the galaxy changes also drastically the
chemical evolution of dIrrs.  If previous episodes of SF carve a very
large cavity, the release of metals from dying stars occurs in a very
hot medium.  Under these conditions, the cooling timescale of the
newly produced metals is very large and, in the presence of a galactic
wind, these metals can be directly carried out of the galaxy (Recchi
et al. 2005; hereafter Paper I).  Consequently, the chemical
enrichment of the galaxy is not influenced by the ongoing production
of metals and the chemical composition only reflects the enrichment
from the first episodes of SF.  However, if the preceding generations
of stars do not provide enough energy to create a galactic wind and to
produce a large cavity of hot gas, the impact of the following episode
of SF can be significant (Recchi et al. 2002).  The same occurs if the
gap between two episodes of SF is large enough to allow the refill of
the center of the galaxy with cold and dense gas.

In this paper, we simulate the evolution of a galaxy after a single
episode of SF of different duration, in order to establish what is the
typical refill timescale when the energy input of SNeIa is
considered.  In Sect. 2 we introduce the model, in Sect. 3 we give our
definition of {\it refill timescale}, in Sect. 4 we present our
results and finally in Sect. 5 some conclusions are drawn.

%__________________________________________________________________

\section{The model}

We simulate a SF region and the late evolution of the hot cavities by
means of a 2-D hydrodynamical code in cylindrical coordinates,
described in detail in Paper I and references therein.  The set-up of
the model is taken from Paper I, namely it reproduces the main
features (\hi mass and distribution, total dynamical mass) of the dIrr
galaxy NGC1569.  In particular, the distribution of gas is
approximately ellipsoidal, with a ratio between minor and major axis
of $\sim$ 0.5, in agreement with observations (Reakes 1980).  Such a
flattened initial distribution of gas favours the development of a
galactic wind in the polar direction, where the pressure gradient is
steepest.  Although the aim of this paper is not to simulate the
detailed chemical and dynamical evolution of this specific object (as
done in Paper I), NGC1569 is a good and well-studied example of a
post-starburst galaxy (Israel 1988), therefore a good benchmark to
study the refill process after an episode of SF.  In order to develop
a physical feeling for the dependence of this process, we test two
possible values of the total \hi mass: $\sim$ 10$^8$ M$_\odot$ (models
labeled ``L'') and $\sim$ 1.8 $\times$ 10$^8$ M$_\odot$ (models
labeled ``H'').  According to the works of Angeretti et al. (2005) we
concentrate on the SF occurring in the central part of the galaxy,
although we do not intend to reproduce in detail the SF history of
NGC1569.  We consider single SF episodes of different duration and
intensity.  We consider either models in which the SF lasts 25 Myr at
a rate of 0.5 M$_\odot$ yr$^{-1}$ (second label ``S''), or models in
which the duration of the SF is 200 Myr at a rate of 0.05 M$_\odot$
yr$^{-1}$ (second label ``L'').  Therefore, for instance, the model
labeled ``LS'' is characterized by a total \hi mass of $\sim$ 10$^8$
M$_\odot$ and a SF rate of 0.5 M$_\odot$ yr$^{-1}$ lasting for 25 Myr.
Model parameters are summarized in Tab.~\ref{table:1}.

The SNeIa rate is calculated according to the so-called
Single-Degenerate scenario (namely C-O white dwarfs in binary systems
that explode after reaching the Chandrasekhar mass because of mass
transfer from a red giant companion).  This rate, in the case of short
SF episodes, peaks after $\sim$ 10$^8$ yr and then declines
proportionally to $\sim$ t$^{-1.8}$ (Greggio \& Renzini 1983;
Matteucci \& Recchi 2001).  The chemical evolution of our models is
also followed, coupling the hydrodynamical simulations with chemical
yields coming from SNeII, SNeIa and intermediate-mass stars.  In
agreement with Paper I, the sets of yields we adopt for massive and
intermediate-mass stars are the ones calculated by Meynet \& Maeder
(2002).  Details on how to trace the chemical enrichment of gas in
this kind of simulations can be found in Paper I and references
therein.

\section{Definition of refill timescale}

According to Paper I, in this work, all the energy and mass input
occurs in the central 200 $\times$ 200 pc$^2$ of the galaxy.  This
assumption comes from the fact that presumably the SF occurring in the
center of NGC1569 comprises the SF of the whole galaxy (see Paper I;
Greggio et al. 1998).  We will therefore define as {\it refill
  timescale} the time elapsed from the end of the SF phase to the
moment at which the physical size of the hot cavity becomes smaller
than the initial 200 $\times$ 200 pc$^2$, namely the time after which,
in spite of the input of energy from SNeIa, a fraction of the central
star forming region has temperatures below 2 $\times$ 10$^4$ K.

\section{Results}

\subsection{A typical evolutionary sequence}

\begin{figure*}[ht]
  \vspace{-6.5cm}
  \centering
% as large as the width of the column
  \includegraphics[width=\textwidth]{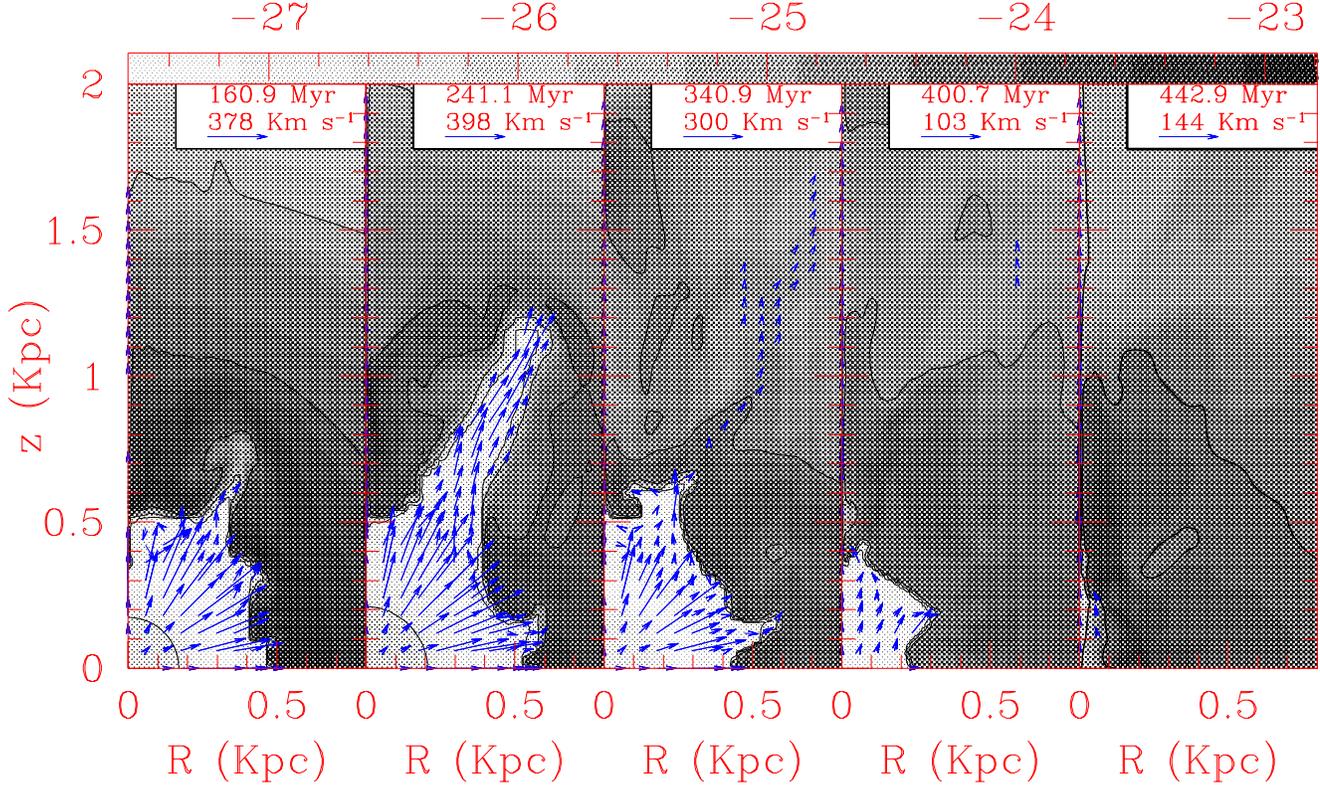}
  \caption{Density contours and velocity fields for model HL 
    (see Tab.~\ref{table:1}) at five different epochs (evolutionary
    times are labelled in the box on the upper right corner of each
    panel).  The logarithmic density scale is given in the strip on
    top of the figure.  In order to avoid confusion, velocities with
    values lower than 1/10 of the maximum value (indicated for each
    panel in the upper right box) are not drawn.  }
  \label{HL}
\end{figure*}

As a representative example of the expected evolution of a galaxy as a
consequence of the energy input of SNeIa, we consider here model HL,
namely the model with a large total \hi mass and an episode of SF
lasting 200 Myr at a rate of 0.05 M$_\odot$ yr$^{-1}$.  Snapshots of
the evolution of this model are shown in Fig.~\ref{HL}.  As one can
discern from the first panel, the combined energy of SNeIa and SNeII
is able to break-out of the galaxy at t $\sim$ 160 Myr and a weak
galactic wind is formed.  The superbubble starts funneling through the
\hi due to the initial stratification of the ISM (see Sect. 2) and to
the structuring of the gas formed by the ongoing dynamical
instabilities.  The energy supply by SNeII ends at t $\sim$ 230 Myr
(200 Myr is the duration of the SF; $\sim$ 30 Myr is the lifetime of a
8 M$_\odot$ star, the smallest able to give rise to a SNII in our
models).  After this time interval, SNeIa still provide enough energy
to sustain the outflow (panel 2).  The funnel begins to shrink at t
$\sim$ 300 Myr and at $\sim$ 340 Myr the outflow has almost completely
disappeared (panel 3).  At $\sim$ 400 Myr the cavity has approximately
the original size of the SF region (panel 4) and from now on it
recedes further towards the center.  The last panel of Fig.~\ref{HL}
shows the density contours a t $\sim$ 440 Myr and at this time the
cavity of hot gas (with temperatures of the order of
10$^5$--10$^{5.5}$ K) has a size of approximately 100 $\times$ 200
pc$^2$; smaller than the SF region.  According to the definition given
in Sect. 3, the refill of the cavity has occurred.  This result might
however depend on the dispersion of SNeIa progenitors (not taken into
account in this work).  Assuming a velocity dispersion of a few km
s$^{-1}$ the progenitors can travel a few hundreds pc after $\sim$ 100
Myr.

The inflow velocity of the cold gas toward the center of the galaxy is
modest (of the order of 5--10 km s$^{-1}$) (approximately the sound
speed of the cold gas), therefore the ram pressure associated with
this flow is of the order of $p_{\rm ram} = \rho_{\rm infall} \cdot
v^2_{\rm infall}$ $\sim$ 10$^{-13}$ erg cm$^{-3}$.  This value is
three to five orders of magnitudes smaller than the ram pressure
associated with the inflows simulated by Tenorio-Tagle \& Munoz-Tunon
(1997).  The inflows considered by these authors should therefore have
a different origin, such as the infall of a high-velocity cloud
directly into the center of a galaxy or a violent large-scale
disturbance in a galactic potential due to a close encounter (Larson
1987).

It is also worth pointing out that the galactic potential well is
dominated by a quasi-isothermal dark halo with a core radius of 1 kpc
(see Paper I for details), therefore the gravitational acceleration in
the center of the galaxy is small.  The infall of cold gas along the
disk is then mainly driven by the pressure gradient originating from
the pressure loss of the cavity once the supershell breaks out (MacLow
\& McCray 1988).  However, one has to emphasize that in these
simulations, at variance with similar studies considering only SNeII
as a source of energy (e.g. DB99), there is no buoyancy of the hot gas
at the end of the SF, since SNeIa provide continuously energy.

\subsection{The evolution of the other models}

\begin{table*}[ht]
%\begin{minipage}[t]{\columnwidth}
\caption{Model parameters and refill timescales}
\label{table:1}
\begin{center}
\begin{tabular}{ccccc}
\hline \hline
Model$^a$ & \hi mass (10$^8$ M$_\odot$) & SF duration (Myr) 
& SF rate (M$_\odot$ yr$^{-1}$) & Refill timescale (Myr) \\    
\hline                        % inserts single horizontal line
   LS & 1.  &  25  & 0.5  & 415  \\      
   LL & 1.  &  200 & 0.05 & 600  \\
   HS & 1.8 &  25  & 0.5  & 125  \\      
   HL & 1.8 &  200 & 0.05 & 200  \\
\hline
\noindent
\end{tabular}
\end{center}
$^a$The identification of a model is made through the notation XY, where X 
indicates the gas mass (``L'' for low and ``H'' for high) and Y 
indicates the duration of the SF episode (``S'' for short and ``L'' for 
long).
%\end{minipage}
\end{table*}

In the last column of Tab.~\ref{table:1} we report of the refill
timescale for each of the considered models.  The refill timescale,
as expected, strongly depends on the total mass of \hi gas at the
beginning of the simulation.  For the extreme model HS (high total \hi
mass, short SF), the refill occurs already $\sim$ 100 Myr after the
SNeII ceased.  The tabulated refill timescales associated with the
``H'' models are consistent with the ones derived for the standard
model studied by DB99, in spite of the fact that no SNeIa were
considered in their simulations. This is due to the fact that their
assumed luminosity in the starburst phase (which lasts 30 Myr in their
models) is $\sim$ 10 times larger than the one adopted in our
simulations.  DB99 also considered a model with a weaker starburst
(their model SB1) and in this case the cavity begins to shrink
immediately after the end of the energy input phase and the cold gas
reaches the origin at t $\sim$ 70 Myr, much earlier than in our
simulations.  The models characterized by a lower \hi mass show larger
refill timescales, which bring the theoretical interval between the
beginning of the SF and the replenishment of the SF region close to 1
Gyr in the case of a prolongated SF (model ``LL'').

\subsection{The impact of a new episode of SF on a refilled cavity}

As demonstrated in Recchi et al. (2004) and in Paper I, if a new
episode of SF occurs in a hot cavity, the metals newly synthesized in
the ongoing starburst are not detectable by the optical spectroscopy
since they are either directly channelled through the galactic funnel,
or they are found in a too hot medium.  This is not the case when the
gap between the two episodes of SF is large enough to allow the refill
of the cavity.  In this case, the starburst injects metals in a much
colder and denser medium, where thermal conduction, thermal
instabilities and eddies allow a fast cooling of the metals and an
effective mixing with the surrounding unpolluted gas.  In order to
show this, we run a model in which, starting from the end of model HS
(see Tab.~\ref{table:1}) we add a second burst of SF, with the same
characteristics as the first one, namely, a SF duration of 25 Myr with
a rate of 0.5 M$_\odot$ yr$^{-1}$.  The onset of this new SF occurs at
$\sim$ 200 Myr, when most of the central 200 $\times$ 200 pc$^2$ of
the galaxy are filled with cold gas.

\begin{figure}
%   \vspace{-1cm}
  \centering
% as large as the width of the column
%  \includegraphics[angle=-90,width=14cm]{long_on.eps}
  \includegraphics[angle=-90,width=\columnwidth]{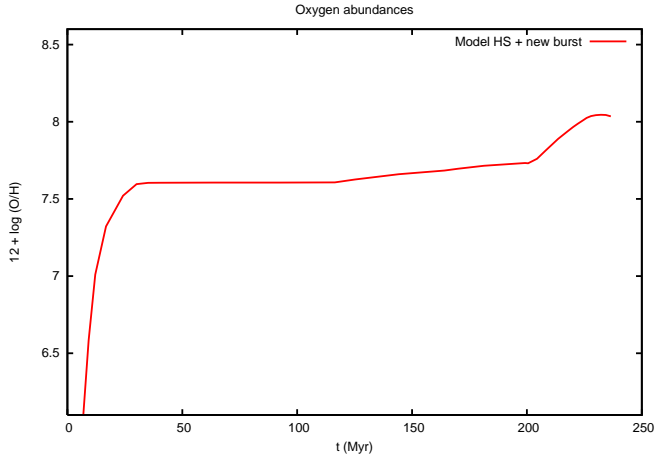}
  \caption{ Evolution of 12 + log (O/H) for model HS 
    (see Tab.~\ref{table:1}) with a second episode of SF at t = 200
    Myr.}
  \label{on}
\end{figure}
   
In Fig.~\ref{on} we plot the evolution of 12 + log (O/H) for this
model, until an evolutionary time of $\sim$ 240 Myr.  As one can see
in this plot, the chemical composition changes significantly $\sim$
10--15 Myr after the onset of the last burst of SF.  The oxygen
composition increases suddenly as a consequence of the input of
freshly produced metals by the ongoing SF.  The mixing timescale of
the metals produced in this last episode of SF is therefore of the
order of 10--15 Myr, in agreement with what is found by Recchi et al.
(2001) in the case of a single instantaneous burst.

\section{Conclusions}

In this paper we have analyzed the late evolution of a galaxy after a
single SF episode in order to explore the timescale needed to refill
the cavity in the center of the galaxy with cold gas once the SF has
ceased.  This timescale strongly depends on the amount of gas
initially present inside the galaxy and ranges between 125 (model with
a large initial amount of \hi and short and intense SF) and 600 Myr
(model with a smaller initial \hi mass and milder and long SF).  This
means that large \hi holes (like the ones observed in many dIrr
galaxies) can survive a few hundred Myr after the last OB stars have
died.  The refill of the cavity is mostly due to the pressure gradient
created after the superbubble breaks out the disk.

A SF occurring in the refilled cavity would produce metals which mix
with the surrounding unpolluted medium in a timescale of the order of
10--15 Myr, at variance with what happens if the center of the galaxy
is still occupied by hot and diluted gas.

\begin{acknowledgements}
  We thank the referee, Prof. Jan Palou\v s for comments and
  suggestions which have improved the clarity of the paper.  S.R.
  acknowledges generous financial support from the Alexander von
  Humboldt Foundation and Deutsche Forschungsgemeinschaft (DFG) under
  grant HE 1487/28-1.
\end{acknowledgements}

\end{document}